\documentclass[iop]{emulateapj} 
\usepackage{natbib}
\usepackage{epstopdf}
\newcommand{\hiir}{H~{\scshape ii}~region}
\newcommand{\hiirs}{H~{\scshape ii}~regions}
\newcommand{\um}{$\mu$m}
\newcommand{\degree}{^{\circ}}

\slugcomment{Submitted to ApJ}

\shorttitle{}
\shortauthors{}

\begin{document}


\title{The Milky Way Project: A statistical study of massive star formation associated with infrared bubbles}


\author{S. Kendrew\altaffilmark{}}
\affil{Max-Planck-Institut f\"{u}r Astronomie, K\"{o}nigstuhl 17, 69117 Heidelberg, Germany}
\email{kendrew@mpia.de}

\author{R. Simpson\altaffilmark{}}
\affil{University of Oxford, Denys Wilkinson Building, Keble Road, Oxford OX1 3RH, UK}

\author{E. Bressert\altaffilmark{1}\altaffilmark{2}}
\affil{School of Physics, University of Exeter, Stocker Road, Exeter EX4 4QL, UK}

\author{M.S. Povich\altaffilmark{3}}
\affil{Department of Astronomy \& Astrophysics, Pennsylvania State University, 525 Davey Laboratory, University Park, PA 16802, USA}  

\author{R. Sherman}
\affil{Department of Astronomy \& Astrophysics, University of Chicago, 5640 S. Ellis Ave., Chicago, IL 60637, USA}

\author{C. J. Lintott\altaffilmark{4}}
\affil{University of Oxford, Denys Wilkinson Building, Keble Road, Oxford OX1 3RH, UK}  

\author{T. P. Robitaille\altaffilmark{}}
\affil{Max-Planck-Institut f\"{u}r Astronomie, K\"{o}nigstuhl 17, 69117 Heidelberg, Germany}

\author{K. Schawinski\altaffilmark{5}}
\affil{Yale Center for Astronomy and Astrophysics, Yale University, P.O. Box 208121, New Haven, CT 06520, USA}

\author{G. Wolf-Chase\altaffilmark{6}}
\affil{Astronomy Department, Adler Planetarium, 1300 S. Lake Shore Drive, Chicago, IL 60605, USA}   

\altaffiltext{1}{ESO, Karl-Schwarzschild-Strasse 2, 87548 Garching, Germany}
\altaffiltext{2}{Harvard-Smithsonian Center for Astrophysics, 60 Garden Street, Cambridge, MA 02138, USA}
\altaffiltext{3}{NSF Astronomy and Astrophysics Postdoctoral Fellow}
\altaffiltext{4}{Astronomy Department, Adler Planetarium, 1300 S. Lake Shore Drive, Chicago, IL 60605, USA}
\altaffiltext{5}{Einstein Fellow}   
\altaffiltext{6}{Department of Astronomy \& Astrophysics, University of Chicago, 5640 S. Ellis Ave., Chicago, IL 60637, USA}




\begin{abstract}
The Milky Way Project citizen science initiative recently increased the number of known infrared bubbles in the inner Galactic plane by an order of magnitude compared to previous studies. We present a detailed statistical analysis of this dataset with the Red MSX Source catalog of massive young stellar sources to investigate the association of these bubbles with massive star formation. We particularly address the question of massive triggered star formation near infrared bubbles. We find a strong positional correlation of massive young stellar objects (MYSOs) and \hiirs~with Milky Way Project bubbles at separations of $<$ 2 bubble radii. As bubble sizes increase, a statistically significant overdensity of massive young sources emerges in the region of the bubble rims, possibly indicating the occurrence of triggered star formation. Based on numbers of bubble-associated RMS sources we find that 67$\pm$3\% of MYSOs and (ultra)compact \hiirs~appear associated with a bubble. We estimate that approximately 22$\pm$2\% of massive young stars may have formed as a result of feedback from expanding \hiirs. Using MYSO-bubble correlations, we serendipitously recovered the location of the recently discovered massive cluster Mercer 81, suggesting the potential of such analyses for discovery of heavily extincted distant clusters. 

\end{abstract}





\section{Introduction}\label{sec:intro}

High-mass stars have a far-reaching effect on a galaxy's interstellar medium (ISM). Throughout their lifetimes their feedback shapes the surrounding cloud material. At the earliest stages of formation, powerful outflows energize the natal molecular cloud, and once switched on, the UV radiation ionizes the young star's surroundings and carves out an H$^{+}$-filled cavity. Ionizing radiation combines with stellar winds to clear out the surrounding gas and dust, forming bright shells, partial or complete bubbles. Upon their deaths as supernovae, massive stars inject as much energy into the ISM as they do during their lifetimes in the form of heating and shocks. These feedback mechanisms combine to shape and disrupt giant molecular clouds (GMCs), and eventually help regulate star formation on a Galactic scale~\citep{Hopkins2011, Matzner2002}.  

\subsection{Infrared bubbles}\label{sec:bubbles}
             
Bright-rimmed bubbles around newly formed massive stars and clusters are readily visible at infrared wavelengths, as the stars' UV radiation excites polycyclic aromatic hydrocarbons (PAHs) in the swept up shell, and hot dust surrounding the young stars radiates at mid-IR wavelengths. Large scale surveys with the Spitzer Space Telescope, such as GLIMPSE at 3.6/4.5/5.8/8.0~\um~\citep{Benjamin2003a} and MIPSGAL at 24/70~\um~\citep{Carey2009a}, and the recently completed all-sky survey of the WISE mission~\citep{Wright2010}, are ideal datasets for identifying such structures. \citet{Churchwell2006, Churchwell2007} (C06, C07 hereafter) made a first attempt at cataloguing bubbles by visual identification in Spitzer/GLIMPSE images, identifying some 600 bubbles over the entire area covered by the survey ($|l| \leq 65\degree$, $|b| \leq 1\degree$). Good correlation of IR bubbles with known \hiirs~and relatively low contamination from supernova remnants (SNR), asymptotic giant branch (AGB) star bubbles and planetary nebulae (PNe) reported in the literature, indicates that bubbles are a useful tracer of star formation activity~\citep{Churchwell2006,deharveng10}. 

The Milky Way Project \citep[S12 hereafter]{simpson_dr1} recently produced an expanded catalog of 5106 infrared bubbles from Spitzer imaging surveys. This order-of-magnitude increase in the number of known bubble objects presents a new opportunity to perform statistical studies of high-mass star formation on a Galactic scale. The sample is likely to be heterogeneous in nature but initial cross-matching with existing catalogs described by S12 indicates that many bubbles are associated with regions of massive star formation.

\subsection{Triggered star formation}\label{sec:triggering}  

Triggered or sequential star formation is the process whereby feedback from energetic events sparks a second generation of star formation. Proposed causes of triggering include supernova explosions~\citep{Phillips2009}; formation of massive stars or clusters~\citep{Elmegreen1977}; massive stellar winds~\citep{Castor1975}; protostellar outflows~\citep{Barsony1998}; and spiral density waves~\citep{Roberts1969} and galaxy-galaxy tidal interactions~\citep{Woods2006} on Galactic scales. If prevalent in galaxies, triggering could provide a mechanism for the propagation of star formation through the galactic ISM, potentially supporting a mode of self-sustaining star formation. Comprehensive reviews of the theory of triggered star formation were recently given by~\citet{Zinnecker2007} and~\citet{elmegreen11}.

Two scenarios for sequential star formation are commonly proposed in the literature. \citet{Elmegreen1977} originally proposed the ``collect and collapse'' process, which occurs when neutral gas becomes swept up and compressed between the expanding shock and ionization fronts from the~\hiir, causing it to fragment and collapse. The theory is further described by~\citet{elmegreen98}. The process acts on large spatial scales along the bubble rim and the fragmentation is thought to result in the formation of massive fragments ($\geq$7 M$_{\odot}$; ~\citealt{whitworth94, dale11a}). The mechanism has been extensively tested, both with simulations~\citep{Dale2007,Dale2009, Gritschneder2009} and observations of shells around known star-forming regions~\citep[e.g.][]{deharveng05, deharveng10, Bik2010a, Brand2011}.

Radiatively driven implosion (RDI) is believed to act on smaller scales when pre-existing condensations inside the clumpy molecular cloud, often in narrow pillars or globules, are compressed and driven to collapse by the pressure from the ionized material. In this case, the timescale for star formation to occur is determined by the ionizing flux and the time taken for the pressure disturbance to reach the condensation~\citep{lefloch94, elmegreen11, miao09, Bisbas2011}. The RDI process has been validated by smoothed particle hydrodynamic (SPH) simulations~\citep{Dale2007a}; hydrodynamical simulations presented by~\citet{Ercolano2012} of ionizing feedback around massive stars are successfully able to reproduce observed morphologies of \hiir~pillars and shells where RDI is thought to take place. \citet{lee07} present observational evidence of RDI near OB associations, finding that the process preferentially produces low- and intermediate-mass stars.

The theoretical picture of triggered star formation is complicated by the fact that feedback from OB stars is known to have destructive effects on further star forming activity in the cloud. Expulsion of the gas by radiation pressure and ionization may halt any ongoing accretion around nearby young stars, preventing the formation of new stars, limiting their mass, or leading to the disruption of the parent cluster~\citep{Boily2003, dale05, Matzner2002, Hopkins2011, Bastian2006}. \citet{Matzner2002} found \hiirs~to be the main source of turbulent energy injection in dense giant molecular clouds, which prevents further protostellar collapse. The effects of ionizing feedback from young clusters on the ISM have also been challenged in the context of triggering by~\citet{Dale2011}. This calls into question the entire scenario of expanding \hiirs~driving massive star formation in the host cloud.

\subsubsection{Observational evidence}
                                                 
The challenge for observational evidence of triggered star formation is establishing the causality between the original and subsequent star formation: did the initial star formation event really cause the following generation, or did the clearing of the cavity simply uncover pre-existing star forming clumps and cores in the cloud material?

The majority of observational studies into triggered or sequential star formation near SNR or \hiirs~take a phenomenological approach, combining multiple datasets (typically a combination of near-, mid- and far-infrared, millimeter and radio wavelengths) to calculate age, mass and luminosity of triggering and triggered sources, and kinematic properties of the young stars and the surrounding ISM. Evidence of triggering has thus been reported near a number of known \hiirs, e.g. Sh2-212~\citep{Deharveng2008}; RCW120~\citep{Zavagno2010}; W51a~\citep{Kang2009}. Such studies offer reasonably convincing evidence of triggered star formation, however frequently conclude with open questions and uncertainties. Furthermore, they cannot address the question of how significant triggered star formation is on Galactic scales.

\subsubsection{A statistical approach: YSO clustering near bubbles}

To address the uncertainties inherent in observations of individual \hiirs,~\citet[T12 hereafter]{thompson12} used a different approach. They performed a correlation analysis of bubble and YSO populations in the inner Galactic plane to investigate YSO clustering properties in the vicinity of IR bubbles. With a well-studied sample of 2000 sources extracted from mid-infrared images (8/12/14/21~$\mu$m), the Red MSX Source (RMS) catalog~\citep{Urquhart2008} provides a suitable dataset of massive young stellar sources for such a study. 

Using the C06 catalog of bubbles from the GLIMPSE survey and a sample of massive YSOs (MYSOs) and ultra-compact \hiirs~(UCHII) from the RMS catalog, they identify a statistically significant overdensity of MYSOs on the scale of 1 bubble R$_{\rm eff}$. If we assume triggering is real, their result allows an estimation of the prevalence of triggered star formation in the inner Galaxy. How much of the Galaxy's star forming activity may have been sparked by preceding star formation episodes?

T12 find that 14\% of their RMS MYSO/UCHII region sample lie within two (bubble) radii from a bubble. Based on their observed overdensity, the authors estimate that the formation of $>$14\% of MYSOs in the inner Galactic plane may have been triggered by feedback from nearby massive young stars or clusters. This however assumes that \emph{all} YSOs found near bubble rims are forming as a result of triggering. Because of the incompleteness of the C06 bubble sample, they present this as a lower limit that may increase by a factor $\sim$2 given a more complete bubble sample. With the first release of bubble data from MWP, such a sample is now available. In this paper we expand the analysis presented in T12 to include the MWP bubbles, to investigate statistically the potential prevalence of triggered star formation on Galactic scales.
                                   
The paper is organised as follows: in Section~\ref{sec:data} we describe the catalogs used for this analysis, their properties and limitations. Section~\ref{sec:method} presents the method used for the correlation analysis, and how it was applied to the datasets. In Section~\ref{sec:results} the analysis results are presented: we first examine the auto-correlation properties of the RMS dataset used in the subsequent analyses. We show our reproduction of the T12 result using the Churchwell/RMS catalogs, and then extend this to the larger MWP sample. We examine correlations for a number of subsamples to investigate the sensitivity of the result to a number of bubble parameters, notably size, thickness, and RMS source type. Limitations of the results and their implications of our findings on the prevalence of triggered star formation are discussed in Section~\ref{sec:discussion}.

\section{Data catalogs}\label{sec:data}
                 
\begin{figure*}
	\includegraphics[width=18cm]{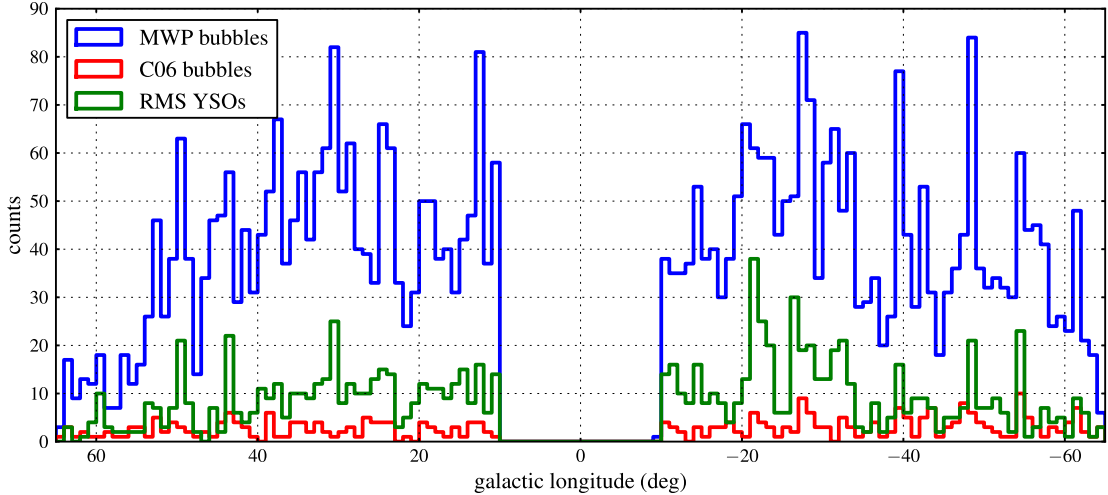}
	\caption{Distribution with galactic longitude of the three catalogs: RMS all young sources (green), C06 bubbles (red) and MWP bubbles (blue). Note that the region $|l|\leq10\degree$ was excluded as this is not covered by the RMS catalog.}\label{fig:londist}
\end{figure*}

\begin{figure}
	\includegraphics[width=10cm]{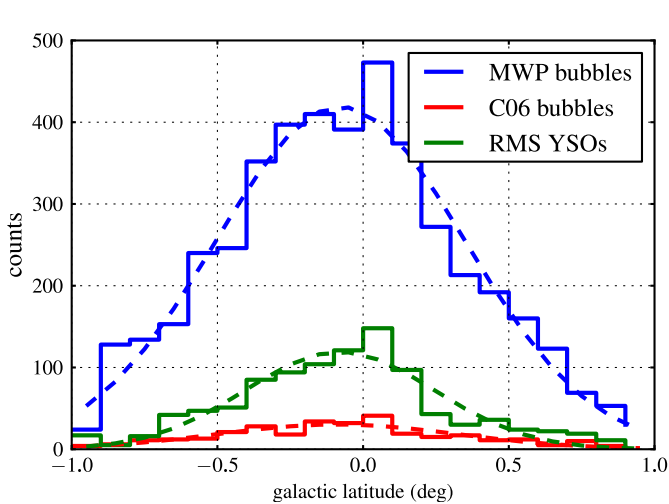}
	\caption{Galactic latitude distribution of the three catalogs: RMS all young sources (green), C06 bubbles (red) and MWP bubbles (blue). Overplotted are the best-fit Gaussian functions, used for the catalog randomisations.}\label{fig:latdist}
\end{figure}

This section describes the catalogs referenced in the work presented in this paper. The region of overlap between the RMS, C06 and MWP catalogs covers $10\degr \leq |l| \leq 65\degr$, $|b| \leq 1\degr$, and for each dataset sources were selected within these limits only. Longitude and latitude distributions are plotted for all three catalogs together in Fig.~\ref{fig:londist} and \ref{fig:latdist}.

\subsection{Milky Way Project bubbles}\label{subsec:mwpdata}

The main focus of the analysis presented in this paper consists of the Milky Way project (MWP) Data Release 1 (DR1) bubbles, presented in S12. These bubbles were identified by over 35,000 users of the MWP citizen science website\footnote{http://www.milkywayproject.org}, a project created by the Zooniverse~\citep{Fortson2008}, in RGB images from the Spitzer Space Telescope GLIMPSE and MIPSGAL surveys~\citep{Benjamin2003a, Carey2009a}. The color composites were created from the 4.5/8.0/24.0\um~images over the coordinate range $|l| \leq 65\degr$, $|b| \leq 1.0\degr$. Images were presented online to users, who were asked to draw the outlines of bubbles using an ellipse-drawing tool. From the inner and outer ellipse sizes, effective radii (R$_{\rm eff}$) and thicknesses (t$_{\rm eff}$) were calculated as simple descriptive metrics for the bubbles, using the equations of C06:

\begin{eqnarray}
	R_{\rm eff}=\frac{(R_{out}r_{out})^{0.5}+(R_{in}r_{in})^{0.5}}{2}\label{eq:reff}\\
	t_{\rm eff}=(R_{out}r_{out})^{0.5}-(R_{in}r_{in})^{0.5} \label{eq:teff}
\end{eqnarray}

\noindent where $R_{in}$, $R_{out}$ are the inner and outer semi-major axes, and $r_{in}$, $r_{out}$ the inner and outer semi-minor axes respectively (C06, S12). The minimum inner and outer diameters of the drawing tool cover 0.45\arcmin~and 0.65\arcmin, respectively. The minimum R$_{\rm eff}$ corresponding to these values, following Equation~\ref{eq:reff}, is 0.27\arcmin. The catalog lists inner and outer diameters (or ellipse axes for non-circular bubbles), eccentricities and position angles. Data catalogs for MWP are publicly available online\footnote{http://www.milkywayproject.org/data}. The site also includes a Data Explorer page that visualises the bubble data.                            

\begin{figure}
	\includegraphics[width=10cm]{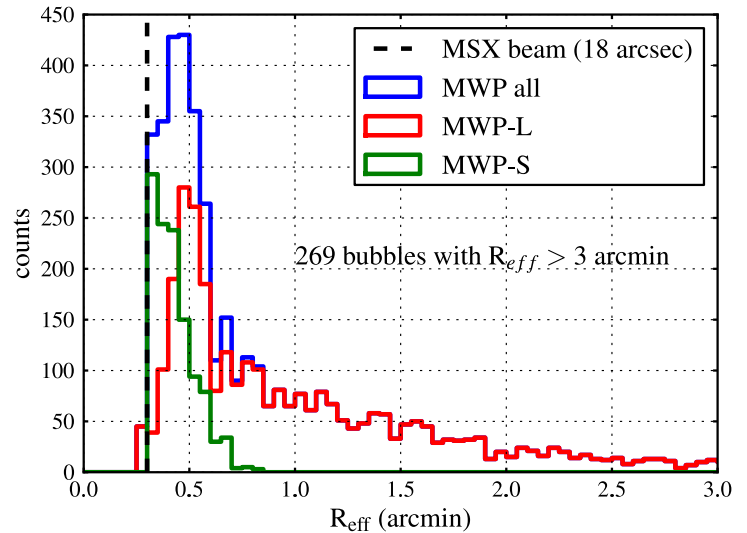}
	\caption{Distribution of MWP bubble effective radii, in arcmin, for the full sample (blue line; 4434 bubbles), MWP-L (red line; 3260 bubbles) and MWP-S (greenline; 1174 bubbles) subsamples. The distribution is truncated at 3\arcmin~ for better legibility of the plot; 269 bubbles have sizes beyond this range. The dashed line indicates the size of the MSX beam.}   \label{fig:mwp_reffdist}
\end{figure}

MWP-DR1 consists of two separate catalogs: the large and small bubbles. The small bubbles were not drawn as ellipses but with simple box shapes. They do not therefore have listed thicknesses in the catalog, nor positional uncertainties. This study uses the combined large and small bubble sample where it overlaps with the region covered by the RMS sources (see Section~\ref{subsec:rmsdata}). The sample contains 4434 bubbles, of which 3260 are `large' and 1174 `small'. Fig.~\ref{fig:mwp_reffdist} shows the distribution of R$_{\rm eff}$ of all bubbles, and of the large and small samples individually. To avoid confusion with simple size descriptions, we identify these samples as the `MWP-L' and `MWP-S' bubbles. 

As noted in S12, several large-scale structural features of the Milky Way galaxy can be traced in the longitude distribution of the bubbles (Fig.~\ref{fig:londist}), most notably the Sagittarius arm near $l=50\degr$, the Scutum arm near $l=25-35\degr$ and $l=-55\degr$, and the Norma arm around $l=-30\degr$ (S12).

\begin{figure}
	\includegraphics[width=10cm]{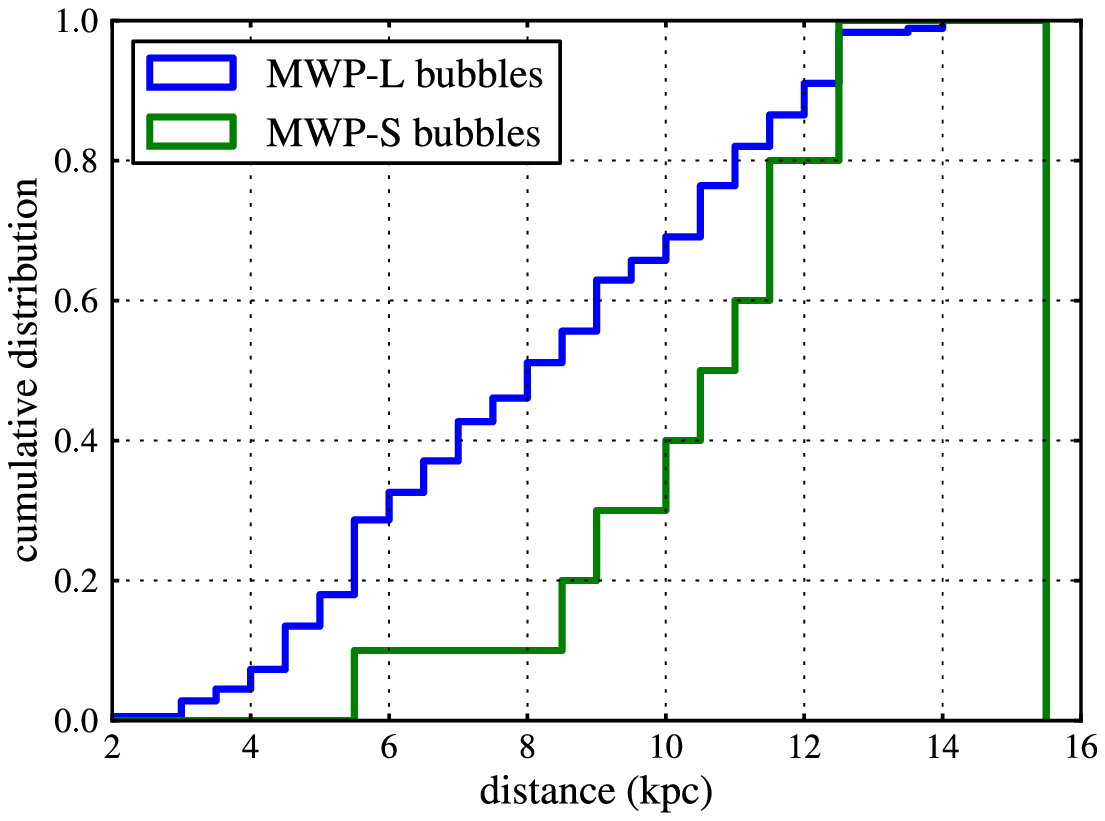}
	\caption{Distribution of known distances to 189 MWP bubbles (179 MWP-L, blue, 10 MWP-S, green) from cross-matching with the~\citet{Anderson2011} catalog of \hiirs.} \label{fig:mwp_dist}
\end{figure}

In S12 the completeness of the catalog was estimated at $>$94\%, based on the decline in bubble discovery rates over time and limitations of the clustering algorithm used to identify bubbles. This limit applies within the size range accessible with the classification tools: the minimum bubble R$_{\rm eff}$ of 0.27\arcmin~determined by the ellipse-drawing tool corresponds to 0.24 pc at 3 kpc, increasing to 1.2 pc at 15 kpc. The size of the largest bubble in the sample measures 11.7\arcmin~in R$_{\rm eff}$. The small bubbles are subject to a higher uncertainty in position and size because of the coarser drawing method used.  Cross-matching with the catalog of \hiirs~of~\citet{Anderson2011} has allowed the distance to be determined to 189 bubbles in the full MWP sample. The distances, presented by S12, cover the range of 2.2 to 14.4 kpc (Fig.~\ref{fig:mwp_dist}). We note that the~\citet{Anderson2011} catalog covers just a small longitude region and contains only sources identified as \hiirs. The distances shown for this small subsample of bubbles may thus not be representative of the full set.



\subsection{Churchwell bubbles}\label{subsec:c06data}
 
C06 described their sample of 322 partial and closed rings visually identified in infrared images from the GLIMPSE survey by a small number of authors, marking the first attempt at cataloguing these objects. They found approximately 25\% of the sample to be coincident with known~\hiirs, and 13\% with open clusters; \citet{deharveng10} report a much higher association, in the range of 85-95\%. Observational biases are described, indicating that bubble discovery can be improved using a larger number of independent visual inspections of the images. Small bubbles in particular are thought to be lacking from the catalog.

In contrast with the Milky Way Project, the Churchwell bubble identifications did not use the MIPSGAL 24~$\mu$m images. We refer to the Churchwell bubbles used in this paper as the `C06' sample, which contains 315 bubbles after limiting the number to the area covered by the MWP sample. C06 make only a rough estimate of the completeness of their catalog (``on the order of 50\% or less''), and indeed the bubble counts in the newer MWP catalog are a factor $\sim$10 higher than C06's sample. The C06 bubble sample is used only for benchmarking of our method used in this analysis, and verification of T12's recent result. Effective radii are defined according to Equation~\ref{eq:reff}.    


\subsection{Red MSX Source catalog}\label{subsec:rmsdata}

The Red MSX Source (RMS) catalog~\citep{Urquhart2008,lumsden02} contains around 2000 sources selected from mid-infrared images from the Midcourse Space Experiment (MSX) Satellite~\citep{Price2001} and near-infrared imaging from the 2MASS survey~\citep{Skrutskie2006}. The sources are detected at a spatial resolution of 18\arcsec. Selection criteria for massive young stellar objects (MYSO), described in detail by~\citet{lumsden02}, are based on photometric properties of known MYSOs. The aim of the survey is to assemble a complete catalog of Galactic MYSOs to $\geq10^4 L_{\odot}$. T12 estimate the sample to be complete to the distance of the furthest bubble in the C06 sample ($\sim$15 kpc; \citealp{deharveng10}). Follow-up work is under way to determine the nature of the sources in the initial catalog~\citep{urquhart11, urquhart12}. Because of confusion and difficulties in distance determination, the region $|l|<10\degr$ is excluded from the catalog.

From the publicly available RMS catalog\footnote{http://www.ast.leeds.ac.uk/RMS/}, we selected those sources collected under the header ``all young sources''. This sample of 1573 sources contains diffuse \hiirs, (compact) \hiirs, YSOs, `HII/YSOs', and a number of sources classified as `young/old star?'; details of these classifications were provided by J. Urquhart (private communication, 2012). The `\hiirs' in the catalog are either compact or ultra-compact, whereas `diffuse \hiirs' are typically extended with respect to the MSX 18\arcsec beam, likely representing a more evolved phase. The `HII/YSO' classification signifies sources that display properties of both object types, perhaps indicating that the source is in a transitional stage, or contains multiple sources in the beam. The `young/old star?' sources display conflicting properties, however, most are thought to be evolved stars. These form a very small portion of our sample ($<$1\%). For simplicity we refer to the sample as `YSOs', however the range of object types this represents in the sample may well be of relevance to the interpretation of the results.

After selecting those sources in the overlap region with the bubble catalogs ($10\leq|l|\leq65\degr$, $|b|\leq 1.0\degr$), 1018 sources remain. Their distribution with galactic longitude and latitude is shown in Fig.~\ref{fig:londist} and~\ref{fig:latdist} respectively. The type classifications by the RMS team suggest the following distribution: $\sim$51\% of sources are \hiirs, $\sim$14\% diffuse \hiirs, $\sim$32\% YSOs and $\sim$2\% HII/YSOs and $\sim$1\% young/old stars.

\begin{figure}
	\includegraphics[width=10cm]{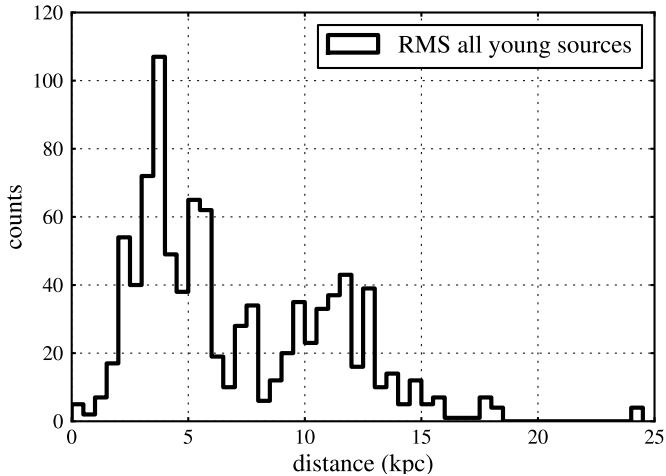}
	\caption{Distribution of distances to the RMS YSOs, for 942 sources with published distances. For those sources with unresolved KDAs (58 sources) the near distance was selected.}\label{fig:rms_distances}
\end{figure}

The RMS catalog contains distances for the majority of sources. These were determined either from the literature or from observations in NH$_3$, CO/CS, methanol or water maser velocity; where the source forms part of a larger complex, the distance to the complex was adopted (from similar measurements or the literature). Out of the 1018 sources in our sample, 74 have no distances in the catalog and for a further 58 the kinematic distance ambiguity (KDA) is unresolved; for this latter group we choose the near distance. The distribution of YSO distances is shown in Fig.~\ref{fig:rms_distances}.


\section{Correlation analyses for clustering studies}\label{sec:method}

Angular correlation functions are a commonly used tool for the identification of e.g. galaxy clusters in dense fields, where they allow for the identification of overdensities in source counts compared with a random distribution~\citep[e.g.][]{Papovich2008,Quadri2008, Hatch2010}. The angular correlation function $w(\theta)$ is defined as the excess probability of finding two objects, or two types of objects, separated by a distance $\theta$. This study employs the commonly used Landy-Szalay estimator~\citep{landyszalay93} for calculating the correlation function $w(\theta)$:

\begin{equation}
	w(\theta)=\frac{N_{DD}-2N_{DR}+N_{RR}}{N_{RR}}
\label{eq:acorr}
\end{equation}

\noindent where $\theta$ is the separation between objects and $N$ represents the pair counts between data points ($N_{DD}$), between data and random points ($N_{DR}$), and between random and random points ($N_{RR}$). Equation~\ref{eq:acorr} applies to one single set of sources, describing thus the auto-correlation or the intrinsic clustering properties of the data. The method can be generalised to be applicable to two different datasets using:

\begin{equation}
	w(\theta)=\frac{N_{D_1D_2}-N_{D_1R_2}-N_{R_1D_2}+N_{R_1R_1}}{N_{R_1R_2}}
\label{eq:xcorr}
\end{equation}

\noindent with the same symbols~\citep{Bradshaw2011}, where subscripts 1 and 2 indicate the bubble and YSO samples respectively. Total pair counts were normalised such that:

\begin{equation}
	\Sigma_{\theta}N_{D_1D_2}(\theta) = \Sigma_{\theta}N_{D_1R_2}(\theta) = \Sigma_{\theta}N_{R_1D_2}(\theta) = \Sigma_{\theta}N_{R_1R_2}(\theta)
\end{equation}

The random bubble catalog was generated with randomly distributed longitudes, latitudes and effective radii. The latitudes were drawn from the best-fit Gaussian latitude distribution (shown in Fig.~\ref{fig:latdist}), and the effective radii follow the best-fit log-normal distribution, to ensure that no artificial over- or underdensities are introduced by the randomisation as compared with the data. The longitudes were distributed uniformly within the coordinate coverage area. Random YSOs were generated similarly in longitude-latitude space. The random catalogs contained a factor of 50 more objects than the corresponding data catalogs, ensuring good sampling of the covered area and sufficient number counts in each $\theta$ bin. 

Bootstrap resampling, implemented via random sampling with replacement of the bubble catalog, was used to estimate sampling errors~\citep{ling86}. 100 bootstrap iterations were carried out for the analysis. Uncertainties are presented at the 1-$\sigma$ level throughout.

Given that massive stars are known to form almost exclusively in clustered environments~\citep{Lada2003a}, a comparison between the bubble-YSO correlation and the YSO auto-correlation is important for the interpretation of the correlation function output. This allows us to assess whether any observed clustering signal is physically meaningful, or simply a reflection of the underlying YSO clustering. To this end we perform an auto-correlation analysis using Equation~\ref{eq:acorr}.

\section{Results}\label{sec:results}

The positional correlation analysis described above was carried out for a number of instances of bubble and YSO catalogs, to compare to the findings of T12 and assess the physical significance of the correlations observed. The code used to perform the analysis was written in Python, and is publicly available (see Section~\ref{sec:code}). 

\begin{figure}
	\centering
		\includegraphics[width=8cm]{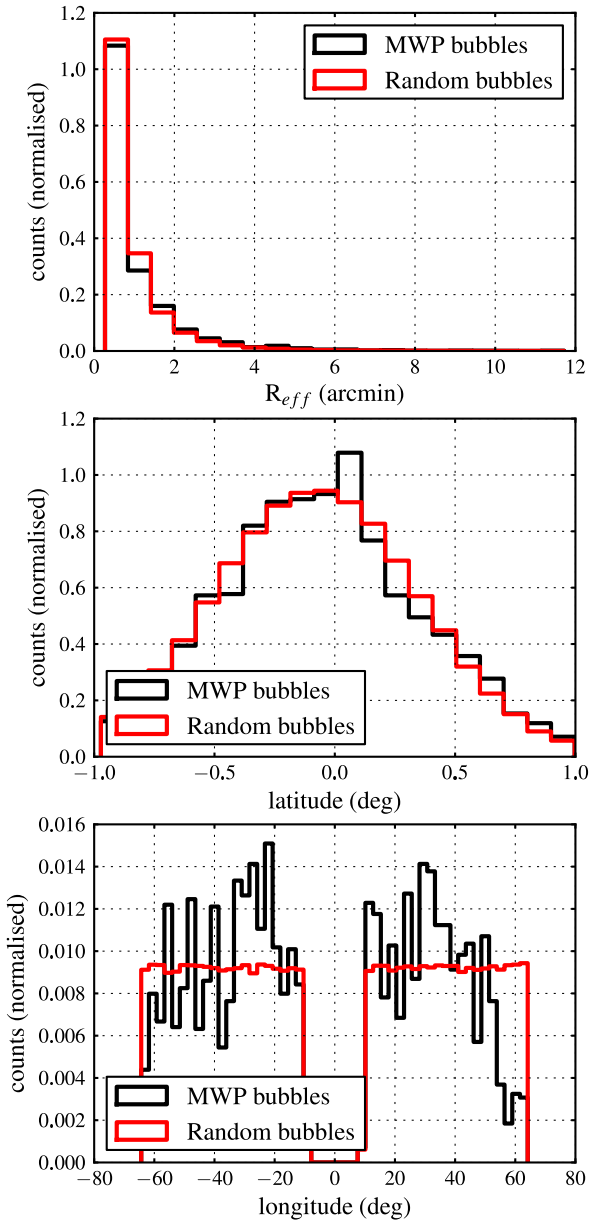}
	\caption{Comparison of distribution with longitude (bottom), latitude (middle) and R$_{\rm eff}$ (top) of the MWP bubbles (black) and their corresponding randomised catalog (red). The random catalog in this case contains 50 times more sources than the data ($\sim2.2\times10^5$ bubbles). The counts are normalised for legibility. The longitude randomisation was uniform over the coordinate range covered; the latitudes and radii follow the best-fit empirical Gaussian and log-normal functions, respectively.}
	\label{fig:data_rand_comp}
\end{figure}

A number of diagnostic plots were produced for each analysis run, to assess the code performance and provide information on sensitivities of the method. Fig.~\ref{fig:data_rand_comp} shows a comparison of the distributions of data and random catalogs for bubbles in longitude, latitude and effective radius. Similar plots were produced to check the YSO random catalog distributions.

\begin{figure}
	\includegraphics[width=10cm]{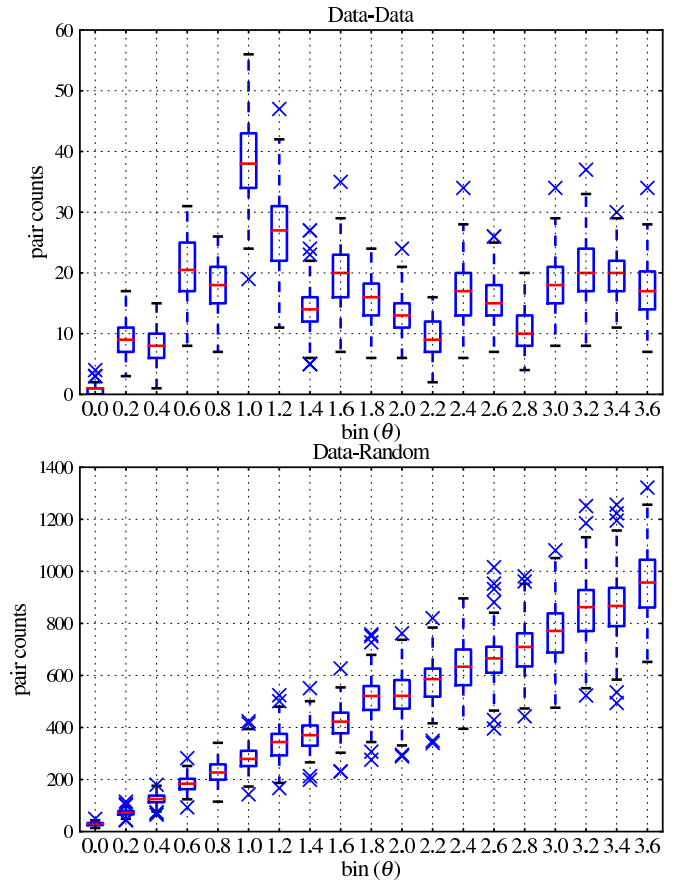}
	\caption{Example of a diagnostic box plot showing the absolute (non-normalised) C06 bubble-RMS YSO pair counts over the bootstrap iterations (N=100) for the computation of N$_{D_1D_2}$ and N$_{D_1R_2}$. The red horizontal lines show the median pair counts in each bin of $\theta$, the boxes span the lower to upper quartiles, and the whiskers show the range 1.5$\times$ the inner quartile range. Outlier points beyond these values are marked with x. Bins are in units of R$_{\rm eff}$, as in the correlation plots.}\label{fig:box_churchrms}
\end{figure}

The second diagnostic plot shown with each analysis is a box plot of the total pair counts, prior to normalisation, in each bin over the bootstrap iterations (e.g. Fig.~\ref{fig:box_churchrms}). The plot shows the median pair counts in each bin of $\theta$ (red horizontal line), the boxes span the lower to upper quartiles, and the whiskers show the range 1.5$\times$ the inner quartile range. Outlier points beyond these values are marked with x. These plots inform about the dispersion and skew of the pair counts across the bootstraps. 
 
\subsection{YSO auto-correlation}\label{subsec:yso_acorr}  

\begin{figure*}
	\includegraphics[width=18cm]{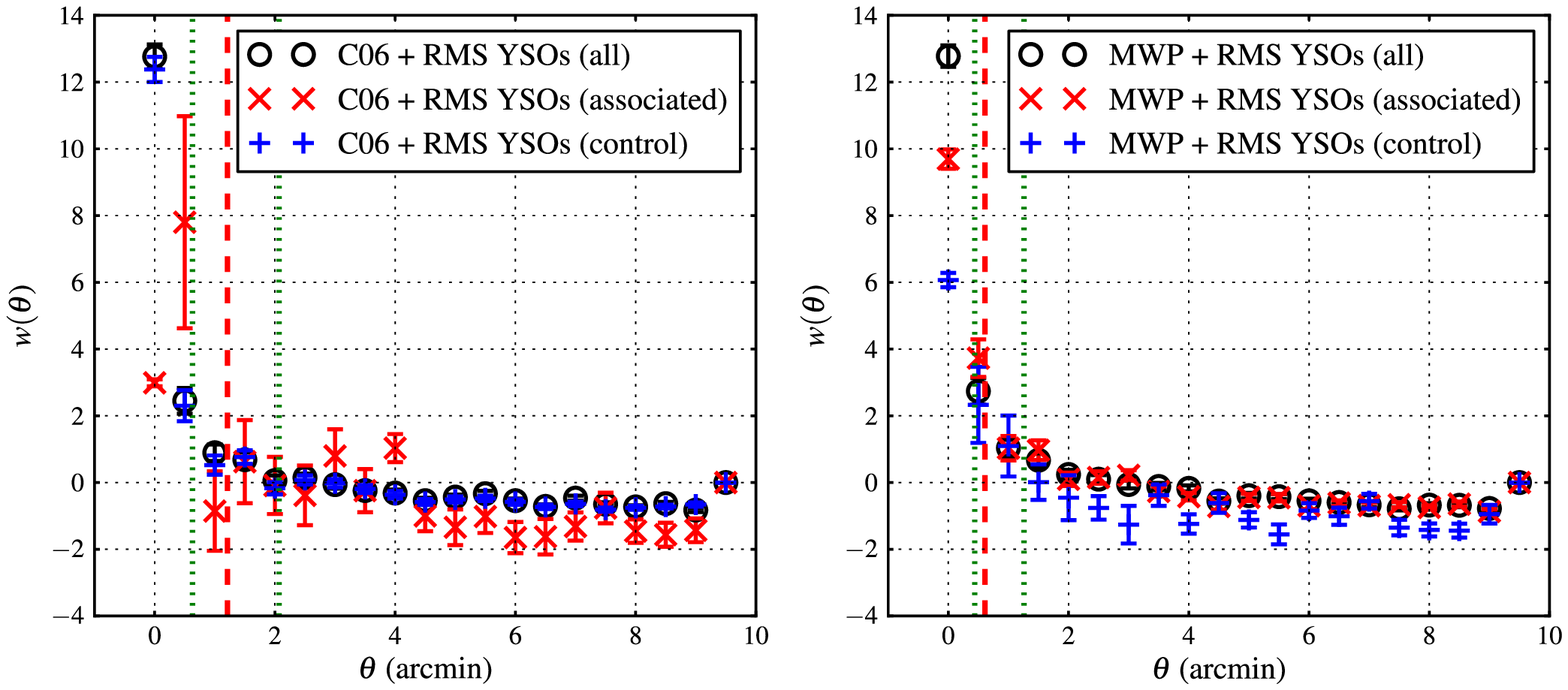}
	\caption{RMS YSO auto-correlation plots. Left: YSO auto-correlation functions for the full sample (circles), bubble-associated (x) and control (+) samples. The dashed line indicates the median R$_{\rm eff}$ of the C06 bubbles, the dotted lines indicate the lower and upper quartile limits of the size distribution. Right: As left, with the sample division and sizes based on the MWP bubbles.}\label{fig:rms_acorr}
\end{figure*}

Important for the interpretation of the correlation plots that follow is the auto-correlation of the YSO sample, which describes the intrinsic clustering properties of these sources. The auto-correlation was calculated using Equation~\ref{eq:acorr}, using a random catalog size containing 50 times the input sample size (roughly 50,000 random sources). Variances were obtained using the bootstrapping method, as described above. We performed 100 bootstrap iterations. 

Following T12, the YSO sample was divided into a `bubble-associated' and a `control' sample. YSO's located within 2 R$_{\rm eff}$ from the nearest bubble are placed in the former group, and those $>$ 3R$_{\rm eff}$ from the nearest bubbles in the latter. Performing this division using both the C06 and MWP bubble catalogs yields striking differences: while the `associated' and `control' groups of YSOs contain 140 and 824 sources (14\% and 81\%) when split using the C06 bubbles, these fractions are very different when using the MWP bubbles. When measured against the full sample of MWP bubbles, 67\% (678) of YSOs lie within 2 radii from a bubble, and the control group contains just 227 sources. 

We can examine this more closely by studying the associations with the MWP-L and MWP-S bubble samples separately. This shows that 644 YSOs lie within 2 radii of a MWP-L bubble, the control group containing 251 YSOs (63\% and 25\% respectively); when comparing with the MWP-S bubbles, only 127 YSOs are `associated' and 865 are `control'. This indicates that the majority of bubble-associated young sources are found in close proximity to \emph{large} (MWP-L) bubbles, which follows naturally from the fact that larger bubbles cover a larger area of sky than their smaller counterparts. It is important to note that this approach cannot distinguish between those RMS sources that are associated with bubbles versus coincident with them; particularly for the MWP-S bubbles, whose radii are comparably-sized to the MSX beam, this is a strong possibility.

By normalising the number of YSOs associated with each bubble by the bubble's area (using the circle traced by 2R$_{\rm eff}$), we find that the mean YSO source density towards MWP-L bubbles is 0.06$\pm$0.23 sources/arcmin$^2$, and towards MWP-S we find a mean density of 0.10$\pm$0.36 sources/arcmin$^2$. When the calculation was repeated using a random catalog of YSOs of the same size, with the randomisation performed as described above, the equivalent mean surface densities towards the bubbles were found to be at least a factor 10 lower than calculated from the real MYSOs, suggesting the YSO surface density enhancement is significant towards \emph{all} bubbles. The variances on the data values are large and the significance of the difference between the MWP-L and MWP-S samples is hard to deduce.  A 2-sample Kolmogorov-Smirnov (K-S) test on the distributions returned a p-value of 0.87, which does not permit us to reject the hypothesis that these two samples were drawn from the same distribution. In other words, we cannot state that the different mean YSO source densities associated with MWP-L and MWP-S bubbles is a statistically significant effect. 

Further insight can be gained from comparing the distribution of source types in the `associated' and `control' YSO samples. Of all diffuse \hiirs~in the sample, 86\% are associated with a bubble; as are 60\% of \hiirs, 65\% of HII/YSOs, 56\% of YSOs and 57\% of young/old stars. The majority of these associations are with MWP-L bubbles, as the above numbers indicate. 

Given the lack of distance determinations for the bubble sample, the physical meaning of these numbers is hard to assess. MWP-S bubbles may be more distant than their large counterparts, in which case the YSO association figures may indicate a completeness limitation in the YSO sample. The increased surface density of YSOs towards MWP-S bubbles may be a result of a higher level of coincidence, rather than `association', between RMS sources and bubbles. Alternatively, if MWP-S bubbles are younger than MWP-L bubbles, star formation may not yet have been triggered, or the central sources may not have swept away enough cloud material to reveal ongoing star formation. A third possibility is that MWP-S bubbles surround relatively nearby stars that are simply not luminous enough to ionize a large bubble and trigger the formation of new stars. Finally, contamination has not yet been studied in detail, and some MWP-S bubbles are likely to be unconnected to star formation (representing instead e.g. PNe or SNR).   

Using the known distances to 189 MWP bubbles, we can compare distances MWP-L (179 bubbles) and MWP-S bubbles (10 bubbles) to assess the likelihood that the two are drawn from the same distribution using a 2-sample K-S test. The test returns a p-value of 0.124, i.e. we cannot reject the hypothesis that the samples are drawn from the same distance distribution. This leaves open the question of the nature of the MWP-S as compared with the MWP-L bubbles, but does not indicate that the MWP-S are likely to be preferentially nearer or further than their MWP-L counterparts.

The auto-correlation was computed for both divisions of the YSO sample. Fig.~\ref{fig:rms_acorr} shows the auto-correlation of the YSOs based on the C06 bubble sample (left) and  the equivalent result using the MWP bubble catalog as a benchmark for the sample division. The clustering properties for the 3 samples are qualitatively very similar, and similar to the auto-correlation when benchmarked against the C06 bubbles. Strong clustering is seen on the smallest scales, with the bubble-associated sample showing a stronger correlation signal than the full and control samples in the range 0.5-1\arcmin~(at 7-$\sigma$). 

The observed auto-correlation properties of the RMS sample suggests that massive young stellar sources are preferentially found in close proximity to each other over what is expected from a random distribution (taking into account typical scale heights); in other words, massive stars, as is known, preferentially form in clustered environments. Importantly, no clustering of YSOs is observed on any spatial scales that are related to bubble sizes; any overdensity observed on characteristic bubble scales is therefore not an intrinsic clustering property of this sample.

\subsection{YSO clustering around C06 bubbles}\label{subsec:church_xcorr}

\begin{figure}
	\includegraphics[width=9cm]{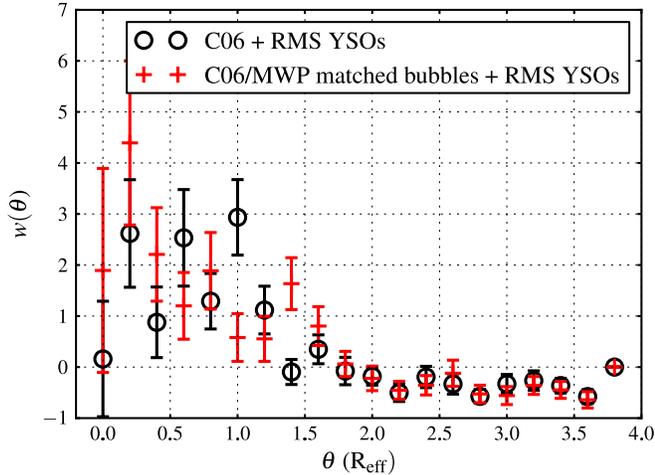}
	\caption{Correlation function between the matched sample of bubbles from MWP-DR1 and C06 (275 bubbles) and RMS YSOs, compared with that of the C06 sample.}\label{fig:corr_matched}
\end{figure}

As a first test the analysis was carried out using the C06 bubble catalog, to verify consistency of the implementation of our method against that of T12. The resulting correlation function is shown in Fig.~\ref{fig:corr_matched} (black circles). A strong positive correlation is observed out to 1 R$_{\rm eff}$, with a peak around 1 $R_{\rm eff}$ observed with a correlation value of $\sim$3 at 4-$\sigma$, confirming the overdensity described in T12. The lower absolute value of the correlation and the somewhat lower significance of the peak can be ascribed to the different binning used in the plot.  Beyond 2 bubble radii, the correlation is effectively non-existent or slightly negative, which would indicate a relative sparsity of sources compared with a random distribution.

The box-and-whisker plots, Fig.~\ref{fig:box_churchrms}, show the median, spread and outliers in pair counts (prior to normalisation) in each bin over the bootstrap iterations for $N_{D_1D_2}$ and $N_{D_1R_2}$, the instances that contain the bootstrapped bubble catalog. As per Section~\ref{sec:method}, sample 1 represents the bubbles (C06 in this case), sample 2 the RMS sources. The plot shows a relatively large spread in pair counts and a clear skewing of the distribution at small separations (bins 1-7), which is a result of the small sample sizes used in this analysis. The low number counts also explain the relatively large error bars in Fig.~\ref{fig:corr_matched}.

A potential source of discrepancy between this result and that of T12 is the different sample of RMS sources used. As described in Section~\ref{sec:data}, we use the publicly available catalog of ``all young sources'' from the RMS survey. This sample contains YSOs as well as diffuse \hiirs~and all evolutionary stages in between. T12 report using a sample of only YSOs and UCHII regions, which contains fewer sources than the set used here. Given the high levels of coincidence between bubbles and diffuse \hiirs, seemingly confirmed by the comparison of bubble-associated and control YSOs, a higher number of \hiirs~in our sample may well increase the overdensity in the YSO counts at the smallest separations.

\subsection{YSO clustering around MWP bubbles}\label{subsec:mwp_xcorr}

The analysis was repeated using the MWP bubble dataset as described in Section~\ref{sec:data}; the YSO catalog was identical to that used in the C06-RMS analysis. As before, random catalogs were constructed with 50 times the number of sources in the input catalog, and 100 bootstrap iterations were performed.

\subsubsection{Analysis checks}

A number of quality checks were performed prior to the full analysis, to examine potential sensitivities and biases of the analysis method to input parameters.

As a first consistency test the analysis was performed with only those MWP bubbles that are also present in the C06 catalog. In S12 the catalogs were cross-matched, and the C06 bubble ID is listed in the MWP data catalogs. This cross-matching between the C06 and MWP catalogs reveals more complex associations than a simple one-to-one matching: in some cases a single C06 bubble contains multiple smaller MWP bubbles, in others multiple C06 bubbles are merged into one large MWP bubble. For this exercise we use only those MWP bubbles that are associated with one single C06 bubble. In the cases where multiple MWP bubbles are associated with the same C06 bubble, we use only the MWP bubble with the closest positional match. Over the relevant coordinate region, this yielded 275 MWP bubbles.

\begin{figure}
	\includegraphics[width=10cm]{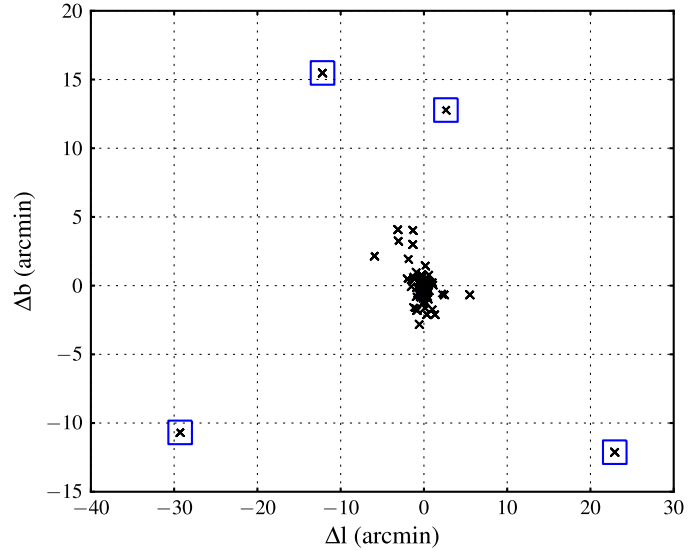}
	\caption{Comparison of position of 275 MWP-C06 matched bubbles. Outliers are marked with squares.}\label{fig:c06mwp_comp}
\end{figure}

\begin{figure}
	\includegraphics[width=10cm]{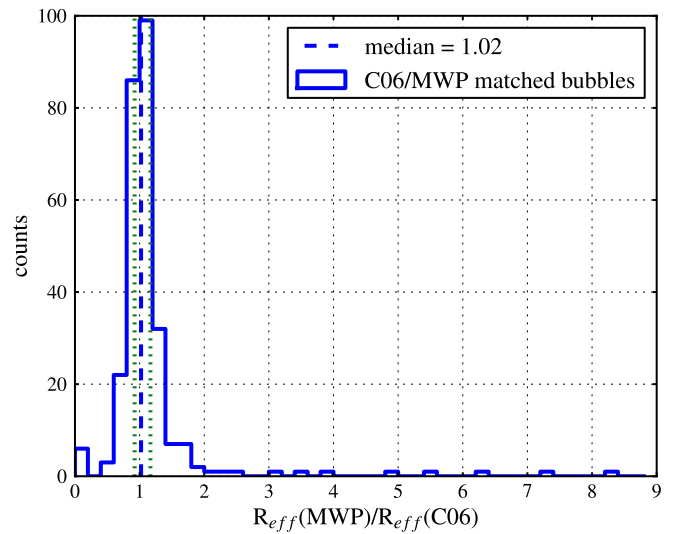}
	\caption{Histogram of the ratio of effective radii of the C06/MWP matched bubbles, showing how the sizes differ for equivalent bubbles in the two catalogs. The dashed line marks the median value of 1.02, the dotted lines mark the limits of the lower and upper quartiles at 0.92 and 1.17, respectively.}\label{fig:c06mwp_reffhist}
\end{figure}

Fig.~\ref{fig:c06mwp_comp} and~\ref{fig:c06mwp_reffhist} show the difference in coordinates and radii of these 275 bubbles. A detailed comparison of bubble parameters in the C06 samples and MWP is also presented in S12.


Median difference in bubble center positions is 0.06\arcmin~in both longitude and latitude. For 90\% of the 275 bubbles the difference is below 0.30\arcmin~and 0.28\arcmin~in longitude and latitude, respectively. Four bubbles, marked with squares in Fig.~\ref{fig:c06mwp_comp}, lie more than 10\arcmin~from their counterparts in C06. Closer inspection reveals their sizes to be discrepant as well, which is likely to be a result of the complexities in the associations described above. The median size ratio of MWP:C06 parameters is 1.02, or 2\%, with the full range covering ratios of 0.034 to 8.3. 90\% of MWP bubbles have sizes within 55\% of the C06 value. These numbers, plotted in Fig.~\ref{fig:c06mwp_reffhist} indicate that the MWP catalog size is typically somewhat larger than that in the C06 catalog for a given bubble. The differences are small compared with the bubble sizes, with some outliers. 

The resulting correlation function is shown in Fig.~\ref{fig:corr_matched}, with the C06-RMS correlation overplotted for comparison. The strong peak at $\theta$=1R$_{\rm eff}$ is not present in the correlation, however two peaks are present at separations of 0.8 and 1.4 R$_{\rm eff}$ with respective significances of 2.5 and 3.2-$\sigma$. The bubbles with large discrepancies between the catalogs in their center coordinates are not thought to affect this result. Those with discrepant radii are likely responsible for the observed differences in the correlation function, as the bubbles' R$_{\rm eff}$ is essentially what the bubble-MYSO separation are measured against. 

The most relevant finding from this test is that the low number counts in the C06 and C06/MWP matched samples make the resulting correlation function very sensitive to small sample difference and choice of binning in $\theta$. Further reducing the numbers by excluding outliers does not improve the significance of the results.

A second question to address is related to the comparative number counts of the datasets used. In the C06-RMS analysis, 315 bubbles and 1018 YSOs are included; a YSO:bubble ratio of roughly 3:1. When repeating the analysis with the full MWP sample of 4434 bubbles, this ratio is reduced to 0.2:1. Can any overdensity of YSOs near bubbles still be recovered when the catalog contains just 1 YSO per 5 bubbles? To investigate this, 4 artificial YSO catalogs were produced with an equal number of sources as the RMS sample, with respectively 100\%, 50\%, 25\% and 10\% of YSOs placed near the rim of a randomly chosen bubble. This was implemented by assigning a radial separation of a random number between 0.8 and 1.6$\times$ R$_{\rm eff}$ of the bubble and a randomly chosen angle between 0 and 2$\pi$. The remaining `fake YSOs' were assigned coordinates of a random `real' RMS YSO. The correlation analysis was carried out between the full MWP bubble sample and each of these fake catalogs, to check whether an overdensity \emph{can} be recovered.

\begin{figure}
	\includegraphics[width=9cm]{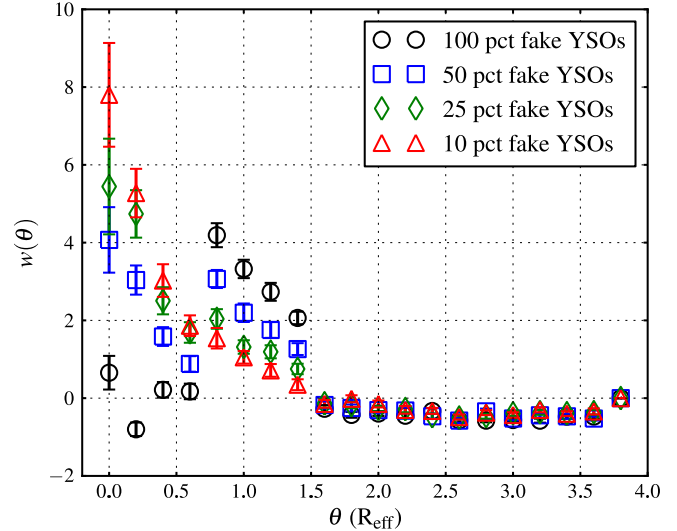}
	\caption{Correlations between MWP-DR1 bubbles and fake YSO catalogs with fractions of YSOs placed deliberately within 0.8-1.6 R$_{\rm eff}$ from a randomly chosen bubble. Data is shown for fake YSO samples containing 100\% (black circles), 50\% (blue squares), 25\% (green diamonds) and 10\% (red triangles) of YSOs on bubble rims, indicating the detection limit of the analysis method.}\label{fig:corr_fakeysos}
\end{figure}

The result of this test is shown in Fig.~\ref{fig:corr_fakeysos}. For each dataset, a positive correlation with high statistical significance is observed in the bins covering $\theta=0.8-1.6R_{\rm eff}$. As the fraction of artificial YSOs in a set decreases, the correlation signal at the smallest bubble-YSO separations increases, and the peak around 1 $R_{\rm eff}$ gradually weakens. The correlation for the `10\% fake YSOs' case is very close to that seen for the full dataset (black circles in Fig.~\ref{fig:mwprms_corr}), as expected, given that this sample contains 90\% of the `real' RMS sources. With increasing `fake' fraction, the correlation at the smallest separations decreases as our test method essentially moves YSOs from bubble centres onto the rim regions. For the extreme case where 100\% of YSOs are `fake', all YSOs are placed near bubble rims, and the correlation at all other separations is roughly zero.

For the sets containing 25, 50 and 100\% of artificially placed YSOs, strong peaks are observed in the bins covering $\theta=0.8-1.6R_{\rm eff}$, indicating that the signal can be recovered even for low number counts of YSOs vs. bubbles, and a substantial fraction of `real' YSOs.

\subsubsection{MWP bubbles-RMS MYSO angular correlation}

\begin{figure}
	\includegraphics[width=9cm]{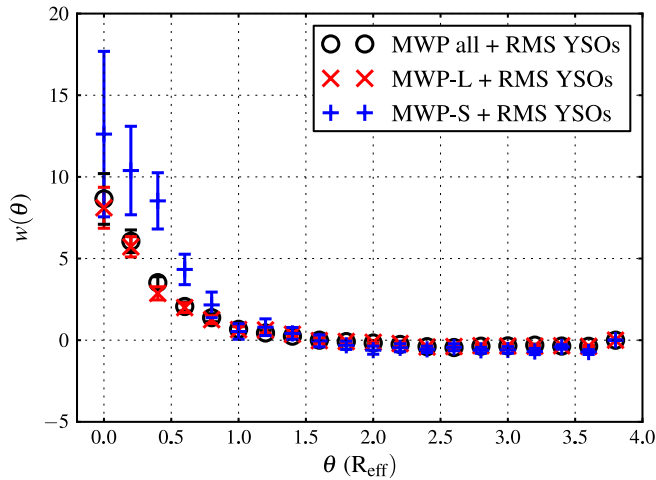}
	\caption{Angular correlation between MWP bubbles and RMS YSOs for the full bubble sample (black circles) and the MWP-L and MWP-S bubble sets (red x and blue + respectively).}\label{fig:mwprms_corr}
   \end{figure}

Following initial quality checks, the analysis was performed for the full MWP bubble and RMS YSO sets. In addition, the correlation was carried out on the MWP-L and MWP-S bubble sets individually. For each case, random catalogs were created with 50 times the number of sources in the input catalog, and 100 bootstrap iterations were performed. The resulting correlations are shown in Fig.~\ref{fig:mwprms_corr}. The improved statistics on the analysis resulting from the larger sample size is reflected in the smaller size of the error bars.

All three correlation functions display a strong clustering signal of the YSOs on scales of $<1 R_{\rm eff}$ of the bubbles in the sample. The correlations of the full sample and the large bubbles are almost identical, displaying a decrease in correlation from 0 to 1 R$_{\rm eff}$. The MWP-S correlation with YSOs appears higher than for the full and MWP-L samples at the smallest separations, consistent with the YSO source density calculations described in Section~\ref{subsec:yso_acorr}.

The correlation function is markedly different from the C06-RMS result presented here and by T12. However, given the huge increase in bubble sample size this should not be unexpected. 

The strong clustering signal in the YSO auto-correlation on small spatial scales, and typical bubble sizes, make it hard to interpret the observed correlation. The analysis check shown in Fig.~\ref{fig:corr_fakeysos} indicates that $>$10\% of $~\sim$1000 YSOs were required to be placed along bubble rims for the signal of the overdensity to become significant over the intrinsic clustering trend of the YSOs. It is therefore not sufficient to look simply at the `associated' and `control' groups in the YSOs, as described in Section~\ref{subsec:yso_acorr}; we need to determine specifically the number of YSOs located near the bubble rims.             

Of the 678 YSOs in the `associated' sample, 225 are located 0.8-1.6 R$_{\rm eff}$ from the center of a MWP bubble. 87\% of these are classified as compact \hiirs~ or MYSOs. The percentage of \hiirs~(55\%) is somewhat higher than in the general RMS YSO sample (51\%), however within the Poisson noise on these number counts this is not a significant difference from the overall YSO type distribution. The analysis check performed with the `fake' YSO catalog suggests that this proportion of YSOs (22\% of the full sample) placed near bubble rims should be recoverable as an overdensity by the correlation analysis. It is possible that the positive correlation is simply `drowned' out by the overdensity at the smallest separations. In the following section, we examine the correlation between subsamples of YSOs and bubbles to see if different correlations are observed.

\subsubsection{Bubble and YSO subsamples}

The large number of bubbles in the MWP sample allows us to explore specific subsamples of bubbles with potentially meaningful properties. The different correlations and YSO source densities for the MWP-S and MWP-L samples, and the difference in correlation functions between the MWP and C06 bubbles, invite a closer examination of the role of bubble size in the bubble-YSO correlation. 

\begin{figure}
	\includegraphics[width=9cm]{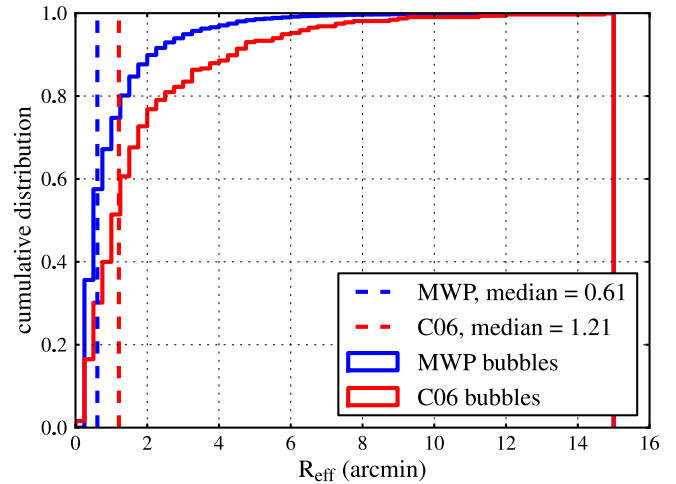}
	\caption{Normalised distribution functions of bubble sizes for the C06 (red) and MWP (blue) samples. Dashed lines indicate the median values, at 1.21\arcmin and 0.61\arcmin for C06 and MWP samples respectively.}\label{fig:c06mwp_reffcomp}
\end{figure}

\begin{figure}
	\includegraphics[width=9cm]{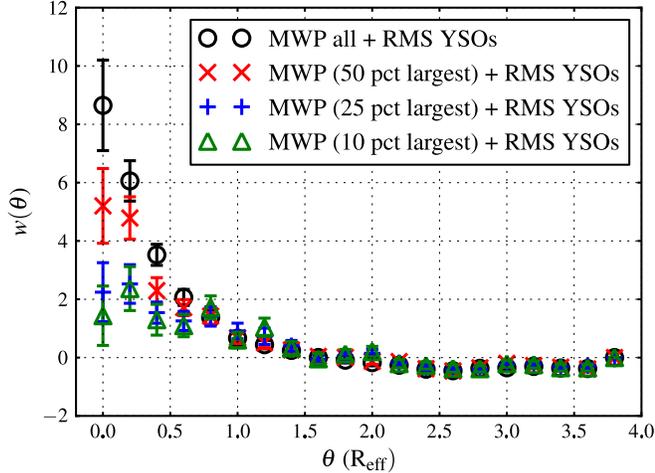}
	\caption{Angular correlation function for the MWP bubbles and RMS YSOs, with the MWP bubbles divided into subsamples based on size, as indicated in the plot legend.}\label{fig:mwprms_corr_sizebins}
\end{figure}

First, the normalised size distribution functions for the two bubble samples are shown in Fig.~\ref{fig:c06mwp_reffcomp}. This clearly shows that the C06 bubbles are large compared with the full MWP set. The dashed lines in the plot indicate the median values of the size distributions; at 1.21\arcmin, the median of the C06 bubbles sizes is twice that of the MWP bubbles (0.61\arcmin). A two-sample K-S test returned a p-value of $\sim$10$^{-4}$, permitting us to rule out simple sampling effects for the observed difference. To investigate the dependency of the correlation function on bubble size, the MWP bubble sample was divided into size bins, containing the 50\% largest bubbles ($>$0.61\arcmin; 2235 bubbles), the 25\% largest bubbles ($>$1.26\arcmin; 1110 bubbles), and the 10\% largest bubbles ($>$ 2.25\arcmin; 448 bubbles). The angular correlation function was calculated for each of these subsamples, the result is shown in Fig.~\ref{fig:mwprms_corr_sizebins}.    

Interestingly, the clustering signal at the smallest separation decreases and a positive correlation around 1R$_{\rm eff}$ emerges as the sample increases in size. The decrease in correlation with increasing size mirrors the lower YSO surface density towards the MWP-L bubbles (Section~\ref{subsec:yso_acorr}). The correlation for the 10\% largest MWP bubbles shows a clear overdensity in the 0.8-1 R$_{\rm eff}$ bin. While the absolute correlation value is relatively low, this point, and that for the 1.2-1.4 R$_{\rm eff}$ bin, carry the highest statistical significance in the series at 4.4 and 3.4-$\sigma$ respectively; these values are very similar to those seen in Fig.~\ref{fig:corr_matched}. 

To rule out that this is simply related to the lower number of sources used in the analysis, we carried out the same correlation analysis with a random selection of 400 MWP bubbles. This was repeated three times, and while the scatter of points in the individual instances can be large, no statistical overdensities appear. We can therefore conclude that the observed overdensity around rims of the largest bubbles is real.

In Section~\ref{subsec:yso_acorr} the YSO surface density was calculated for the MWP-S and MWP-L samples. To investigate the overdensity observed along the rim of the bubbles, the same calculation was performed for those YSOs lying between 0.8 and 1.6 R$_{\rm eff}$ only - the area associated with the rim - and compared with the mean surface density towards the entire bubble. The mean YSO surface density towards the 10\% largest bubbles (within 2 R$_{\rm eff}$) is very low at 0.01$\pm$0.022 sources/arcmin$^2$ over a 2R$_{\rm eff}$ area, consistent with Fig.~\ref{fig:mwprms_corr_sizebins}. However for the MWP-S bubbles, just 37\% of associated YSOs are found in the region around the rim when normalised to the respective area, whereas for the 10\% largest MWP bubbles this value is 53\%. In summary, larger bubbles are generally associated with fewer YSOs than their small counterparts once their larger projected area is taken into account, but their associated YSO population is more likely to be found in the shells rather than in their interiors.

Both the C06 and S12 authors identified a strong correlation between bubble effective radii and effective thicknesses. Given the overdensity of YSOs near the rims of the largest of bubbles, we intuitively expect a similar observation for the angular correlation between YSOs and the thickest MWP bubbles. The correlation analysis was similarly calculated for the MWP-L bubbles, and as expected the correlation function mirrors that shown in Fig.~\ref{fig:mwprms_corr_sizebins}.

\begin{figure}
	\includegraphics[width=9cm]{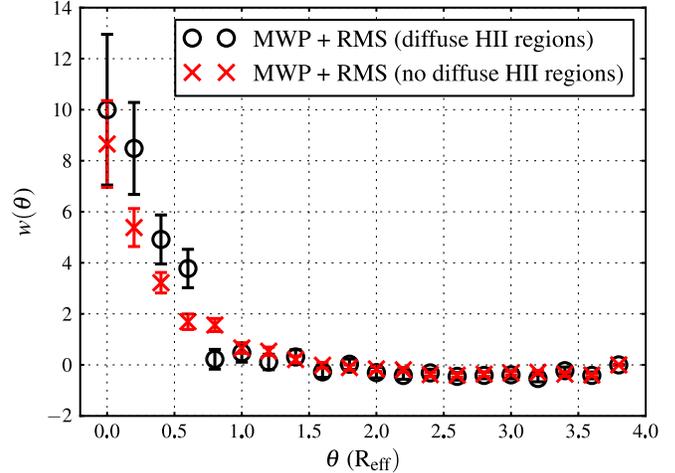}
	\caption{Angular correlation function of MWP bubbles and RMS YSOs, divided into diffuse \hiirs~only (black circles) and other source types (red x).}\label{fig:mwprms_nodiff}
\end{figure}

In a final examination of subsamples, we divided the YSO sample into the diffuse \hiirs~and the other source types, and studied the correlation between bubbles and these subsamples separately. As many bubbles are known to enclose evolved \hiirs, studying these sources separately  may reduce the strong overdensity seen at the smallest YSO-bubbles separations, possibly making an overdensity near the bubble rims more prominent. This would also support the idea that the diffuse \hiirs~are (in part) responsible for the strong central correlation between bubbles and RMS sources, as they trace the bubble driving sources rather than secondary star formation. The result is shown in Fig.~\ref{fig:mwprms_nodiff}. As expected, the correlation at the smallest separations is reduced for the sample excluding diffuse \hiirs. The diffuse \hiirs~ show a stronger correlation with bubble center positions than the other source types, with values reaching zero correlation  before 1R$_{\rm eff}$, while the correlation in the rest of the sample remains positive at this point. In the correlation between bubbles and the sources without diffuse \hiirs, the 0.8R$_{\rm eff}$ data point is reminiscent of the same point seen for the ``10\% fake YSO'' correlation in Fig.~\ref{fig:corr_fakeysos}, hinting at a potential shell-associated overdensity, but this is uncertain. The result of Fig.~\ref{fig:mwprms_nodiff} supports the idea that the diffuse \hiirs~in the sample are more likely to be associated with the trigger\emph{ing} rather than the trigger\emph{ed} sources. 

\section{Discussion}\label{sec:discussion}

We discuss here the physical implications of the findings presented above. We first describe possible limitations and biases in our datasets and methods, and discuss the implications for the occurrence and prevalence of triggered star formation on Galactic scales. Finally, we highlight a serendipitous discovery made by studying particular instances of bubble-YSO correlations.

\subsection{Limitations of the data and methods}\label{subsec:limitations}
                           
To place the findings presented here into context, we must consider first the limitations of the data and methods used in the analysis. The following aspects are of particular relevance: completeness; astrometric precision and spatial resolution; source type uncertainties. These limitations are described here for both RMS and MWP catalogs.

The completeness of the RMS source catalog is seemingly well understood: T12 estimates the sample to be complete for sources with L$>10^4$L$_{\odot}$ over the distance covered by the C06 bubble sample, which is estimated to be $\sim$15 kpc~\citep{deharveng10}. Without distance determinations for a significant fraction of the MWP bubble sample, however, it is unknown whether the distance range covered is similar to that of C06. The 189 MWP bubbles that have assigned distances do cover a similar range, from 2 to 15 kpc approximately. The MWP bubble sample itself is limited in completeness by the size limitation of the drawing tools, as described in Section~\ref{sec:data}, and some bubbles might have been lost through imperfections in the clustering algorithm used to reduce the classification data into the catalog.

Important uncertainties in the results presented here stem from the spatial resolution of the RMS sources, particularly in relation to the typical bubble sizes. The RMS sources are detected at 18\arcsec~in all bands. Given a median bubble radius in our full MWP sample of 0.61\arcmin, and a minimum bubble R$_{\rm eff}$ of 0.27\arcmin, RMS sources are comparably-sized to the smallest bubbles; even median-sized bubbles are only a factor of $\sim$3 larger than the RMS sources. Thus the uncertainty on a source's placement on a bubble rim should be considered to be large, although random differences are expected to be averaged out. This also means we cannot account for those RMS sources that are coincident with bubbles as opposed to associated with them (e.g. on the rims). 

A second related point in relation to the RMS sources is the possible association of diffuse and compact \hiirs. Compact \hiirs~may not necessarily be separate entities from their diffuse counterparts, marking for example density peaks of an extended ionized region. \citet{Morgan2004} showed how bright-rimmed clouds surrounding diffuse \hiirs~can appear as compact \hiirs. Studies of individual sources should be able to distinguish between RMS objects representing triggering versus triggered sources (and thus the objects' evolutionary stage) with relative ease; particularly for those regions for which radio data are available.

Sizes from the MWP-S bubble sample, in addition, should be taken with caution: the drawing method for these bubbles uses a simple box-drawing tool rather than the more precise ellipse-drawing tool used for the MWP-L bubbles. Association of the MWP-S bubbles with RMS sources in particular are expected to be somewhat more uncertain as a result. 

The astrometric precision of the RMS sources is reported by ~\citet{lumsden02} to be 2\arcsec, however these authors found that astrometry of a small but non-negligible number of sources was very poor. Similarly,~\citet{Mottram2007} compared RMS astrometry with mid-IR higher resolution imaging at 10~$\mu$m, and found a 1-$\sigma$ positional offset of 2\arcsec, but rising to 5-10\arcsec~in more complex regions. The positional uncertainty on the MWP bubbles was determined from the spread of center coordinates in the input classifications, for the MWP-L bubbles only (the ``dispersion'' column in the data catalogs). The median ratio of positional dispersion to R$_{\rm eff}$ for these bubbles is 0.3. Combining this value with a worst-case astrometric precision of 10\arcsec~for the RMS sources, the resulting uncertainty on bubble-YSO separation could be as high as 0.7 R$_{\rm eff}$ for the smallest bubble, 0.4 R$_{\rm eff}$ for the median bubble and 0.3 R$_{\rm eff}$ for the largest bubble in the MWP-L sample. Random errors in the separation are expected to average out over a large sample, however we still expect this to limit the significance of our results.

Finally, the analysis uses the effective radius as an indicator of the location of the bubble rim. This metric gives a convenient approximation for the size of a bubble but ignores the bubble's eccentricity, rim thickness, possible breaks or blowouts in the shell, and any other complexities. In the second phase of the Milky Way Project, launched in March 2012, images of catalogued bubbles will be presented to users for the purpose of gathering more precise position and size information for future data releases from the project.

\subsection{Interpreting correlation functions in three-dimensional space}

The search for overdensities of MYSOs and \hiirs~ along bright bubble rims is based on empirical evidence of star formation occurring in these regions~\citep[e.g.][]{Deharveng2008, Zavagno2010, Kang2009}. Bubble morphologies are however three-dimensional (3D) and often very complex given their dense and turbulent surroundings. Projection effects and ellipsoidal morphologies can place driving sources of the bubble expansion in apparently off-center locations, and the velocity dispersions of MYSOs may further complicate the observed bubble-YSO association. The difficulty in studying the 3D geometry of bubbles was shown by~\citet{Beaumont2010}, who used CO observations to recover a flattened ring-like rather than spheroidal morphology for a sample of 43 C06 bubbles.

A detailed interpretation of the correlation functions presented here requires a thorough theoretical investigation of the expected appearance of bubbles and their associated YSOs in the presence of triggering, that takes into account 3D morphologies, projection effects in the Galactic plane, and realistic YSO velocity dispersions. Such models can be used to reconstruct the expected shape of the correlation functions. Such a study will form the subject of a follow-up paper.

\subsection{On the prevalence of triggered massive star formation}
                                
The analysis presented here shows that massive young stellar objects and \hiirs~show a strong positional correlation with the infrared bubbles that populate the galactic ISM. The YSO surface density is greatest towards smaller (thinner) bubbles, consistent with the recent findings by T12. The correlation at the smallest bubble-MYSO separations decreases with increasing bubble size. 

The increased correlation between YSOs and bubbles at the smallest separations is most likely due to a higher level of completeness of the MWP sample, and as a result more RMS ultra-compact, compact and diffuse \hiirs~are found to be associated with an infrared bubble. In addition, many RMS sources located near bubble centers are likely to trace the bubble driving source rather than second-generation star formation. Driving sources of large (nearer or more evolved) bubbles appear extended in the MSX beam and are therefore less likely to be included in the point source catalog. The diffuse \hiirs~included in the RMS catalog are clearly strongly centrally associated with MWP bubbles (see Fig.~\ref{fig:mwprms_nodiff}), but we note that (ultra-) compact \hiirs~seen in RMS may represent the bright-rimmed clouds often associated with their larger diffuse counterparts.

The positional uncertainties described above and the geometric approximation to the bubble rims does not allow for a more precise positional correlation. This is a particular problem for the MWP-S bubbles, whose positional uncertainties cover a significant fraction of their radii. A more sophisticated correlation that makes use of the full ellipticity information of the bubbles could refine the result, particularly using a future data release from MWP, which will include better estimates of uncertainties on the bubble parameters and improved statistics from an increased number of classifications.

A significant overdensity of MYSOs is observed towards the rims of the largest bubbles. Numerous observers have found evidence of ongoing star formation, both low- and high-mass, around bubbles and and shells in known star forming regions; many interpret this as evidence of triggered or sequential star formation. What do our findings tell us about the occurrence or the prevalence of triggered star formation?  

As the bubble radius is a dynamical quantity (i.e. changeable over time and strongly environment-dependent), it is hard to conceive of a selection effect that produces the observed overdensity of YSOs near the rims of large bubbles. 

Following criteria set out by~\citet{elmegreen11} for the identification of triggered star formation, we conclude that the analysis performed here can neither confirm nor reject conclusively the presence of triggered star formation in the Galaxy. The result is strongly limited by the lack of distances for the MWP (and C06) bubbles, which do not allow a firm determination of bubbles' physical sizes or ages. A firm age determination furthermore is non-trivial, as it requires knowledge of the luminosity and spectral type of the driving source(s) of the bubble expansion, and information on the ambient ISM density. We are therefore limited to a qualitative interpretation of the results.

The observed overdensity of RMS sources projected towards the rims of the largest bubbles appears consistent with a mode of triggered star formation. \citet{whitworth94} and~\citet{Dale2007} calculate timescales for shell fragmentation around expanding \hiirs, possibly leading to collect and collapse-driven star formation: for a bubble blown by a single late-type O star, expanding into a 100~$cm^{-3}$ uniform neutral medium, shell fragmentation will typically occur after $\sim$3 Myr. Typical ages of the RMS sources in our sample are of the order of $10^4-10^5$ years (the diffuse \hiirs~are more evolved). Although many parameters come into play, it seems reasonable to assume that only the larger of the MWP bubbles are likely to have a sufficient age, and are driven by sufficiently energetic sources, for a second generation of star formation to have occurred. 

T12 conclude that the majority of MYSOs they find within 2 radii from a C06 bubble,$\sim$14\% of the total sample, are likely to be triggered. Overall we find that 67$\pm$3\% of RMS sources are located within a projected distance of 2 bubble radii, and 22$\pm$2\% are found in the region of 0.8-1.6 R$_{\rm eff}$ (assuming Poisson statistics). Following T12's definition of MYSO's `associated' with bubbles, we could state that 67\% of MYSO may have been triggered, however given the uncertainties and the results with bubble subsamples we posit that the lower figure is more accurate.

As for the \emph{general} prevalence of triggered star formation, our analysis is further limited by the choice of the RMS catalog of YSOs. The large beam of the MSX satellite (18\arcsec) means that a significant fraction of the sources are likely to contain multiple YSOs, particularly those at large distances (18\arcsec at 15 kpc corresponds to $>$1 pc in size). In one follow-up study using 10~$\mu$m ground-based imaging, \citet{Mottram2007} found a multiplicity of 25\% in a sample of $\sim$350 RMS MYSOs. 

While the RMS catalog may contain contributions from low- and intermediate mass young stellar populations, this regime is not well covered by the catalog. As RDI is thought to preferentially produce low- and intermediate-mass stars~\citep{lee07}, our method may not be sensitive to all triggered populations near the MWP bubbles. \citet{whitworth94} showed that gravitational collapse of a dense shell surrounding expanding shells, thought to lead to collect \& collapse star formation, produces predominantly massive fragments. \citet{dale11a} however argue that the fragment mass function cannot be reliably used to predict the mass function of resulting stars or clusters. Our use of the RMS catalog therefore gives a slight bias towards detecting triggered star formation from the collect \& collapse mechanism but our results cannot convincingly distinguish between modes of triggering.


This study highlights that, although challenging to produce, robust and well-documented catalogs of young stellar sources are extremely important for studying star formation on a Galactic scale using statistical methods.

\subsection{A distant massive cluster rediscovered using bubble-YSO associations}\label{subsec:bubprops}

Of all bubbles associated with RMS YSOs, the median number of associated RMS sources is 1. Given the number counts in each dataset, this low association figure is unsurprising. More than 20 MWP bubbles are however associated with at least 5 RMS sources, and our highest-association bubble offers an interesting case study.

\begin{figure*}
	\includegraphics[width=18cm]{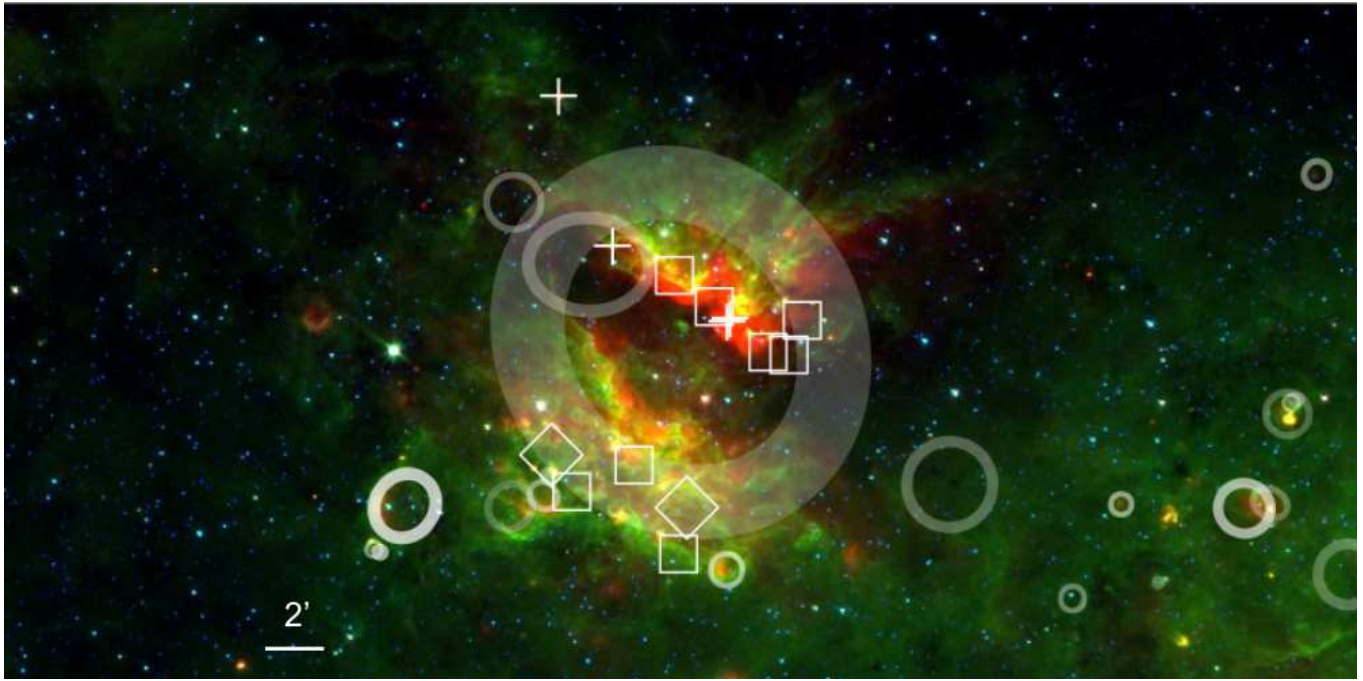}
	\caption{Bubble MWP1G338393+001277, the MWP bubble with the highest number of associated RMS young stellar sources within 2 radii (R$_{\rm eff}$=6\arcmin) from its center, which was found to enclose the newly discovered massive cluster Mercer 81 at a distance of 11$\pm$2 kpc~\citep{Davies2012}. The image was created from 4.5/8.0/24.0\micron~images from the Spitzer GLIMPSE/MIPSGAL surveys. Additional MWP-L bubbles in the field are also shown as ellipses. RMS sources are shown with + (YSOs), squares (\hiirs) and diamonds (diffuse~\hiirs)}.\label{fig:mercer81}
\end{figure*}

The MWP bubble with the highest number of associated RMS sources, shown in Fig.~\ref{fig:mercer81}, is located at $(l,b)=(-21.6, +0.13)$. Bubble MWP1G338393+001277 measures approximately 6\arcmin~in diameter and has a relatively high eccentricity of 0.55. It is surrounded by several smaller bubbles in a hierarchy and 14 RMS young stellar sources, indicating that star forming activity is likely to be ongoing in the region. The bubble can be identified as the shell surrounding Mercer 81, the recently discovered massive stellar cluster at the far end of the galactic Bar near the intersection with the Norma arm~\citep{Davies2012}. \citet{Davies2012} place the cluster at 11$\pm$2 kpc, giving the main bubble a physical radius of $\sim$20 pc. The distance determination is consistent with the RMS catalog distances, which place 12 of the 14 sources projected towards the bubble at 12.9 kpc. With an estimated mass of $\geq 10^4 M_{\odot}$, Mercer 81 is one of just 2 known young clusters at the far end of the Bar. 

The environment of Mercer 81 demonstrates the problem of morphological complexity discussed in Section~\ref{subsec:limitations} in the context of our analysis very well: the bubble is clearly eccentric and opened out to the South. Even with the ellipse-drawing tool the bubble shape is only approximately matched to the actual morphology. The RMS sources clearly trace the bright PAH emission along the rim, but this region is not very well described by the user-drawn ellipses.

Given a visual extinction towards the cluster of $A_V\sim45\pm15$ and the heavy stellar crowding at optical and NIR wavelengths, such clusters are extremely challenging to detect using traditional search methods. Our serendipitous recovery of this cluster based purely on associations of bubbles and young stellar sources demonstrates the potential of the MWP dataset to explore not just massive star formation but potentially discover heavily extincted clusters to large distance in the Galactic plane.   

The discovery of the massive cluster powering \hiir~RCW 79 based solely on the presence of an \hiir~with star formation occurring in its vicinity, reported in \citet{martins10}, offers another noted example.

\section{Conclusion}

We studied the statistics of massive star formation in the vicinity of infrared bubbles in the inner Galactic plane, recently catalogued by the Milky Way Project, a Zooniverse citizen science initiative. Our detailed statistical correlation analysis between these bubbles and the RMS catalog of MYSOs and \hiirs~shows a high level of clustering of MYSOs and \hiirs~and the bubbles, reflecting the clustered mode of massive star formation on the one hand, and indicating a strong association of bubbles with massive star formation activity on the other. 

The largest (and thickest) bubbles in the MWP sample show a statistically significant overdensity of young stellar sources near their rims. A qualitative analysis of timescales for bubble expansion and shell fragmentation indicates that these large bubbles are likely to be old enough to have caused the collapse of their dense shells. While our analysis cannot prove or disprove the occurrence of triggered star formation in these regions, the finding supports a collect \& collapse-driven mode of triggered star formation on the rim of expanding bubbles. We find that 67$\pm$3\% of the RMS young stellar sources lie within 2 radii of a MWP bubble, and 22$\pm$2\% lie within 0.8-1.6 radii.

The most important caveats of this result are related to the uncertainty on the bubble-YSO separation, due to the astrometric precision and extended sizes of the sources in both catalogs, the geometric approximation used for the bubble sizes, and the uncertainty in source types for the RMS sources. Thus we are unable to distinguish between those RMS sources that are coincident with bubbles versus those that are potentially triggered nearby. The interpretation of the results is limited by the lack of MWP bubble distances and the complexity of characterising bubble physical sizes and ages.

Whereas many studies of triggered star formation have focused on detection of star formation activity in known star forming regions, the analysis presented here shows how statistical methods can be used to identify potential regions of triggering in an unbiased way. The independent recovery of a recently discovered massive cluster at 11$\pm$2 kpc with $A_V\sim45\pm15$, Mercer 81, by closer investigation of a strong bubble-MYSO association alone demonstrates the potential of statistical comparisons of large-scale multi-wavelength Galactic surveys - not just for the study of Galactic-scale star formation, but for the discovery of stellar clusters to large distances.

The advent of data from large-scale Galactic surveys at infrared, submillimeter and millimeter wavelengths (e.g. WISE, \citealt{Wright2010}; ATLASGAL, \citealt{Schuller2009}; and HiGAL, \citealt{Molinari2010a}) offer the exciting prospect of extending such statistical studies to longer wavelengths and larger areas of the Galaxy, allowing for a more complete census of Galactic star formation and a more comprehensive view on the lifecycle of molecular material in the Milky Way. This was recently illustrated using ATLASGAL survey data by~\citet{Beuther2012}.

Finally, it is important to note that the ability, provided by the MWP's large catalog, to constrain the correlation function of bubbles provides a stringent test for future models of triggered star formation. Given appropriate assumptions or measurements of the age of a population of YSOs and young stars a model of triggering that includes estimates of efficiency and an expansion of the bubbles should predict the form of the observed correlation function.    
 
\section{Code}\label{sec:code}
                                    
The main body of code to perform the correlation and auto-correlation analyses presented in this paper was written in Python version 2.6, and is available online via the journal. In addition, it is freely available for download as a public Github repository webpage\footnote{https://github.com/skendrew}. We invite and encourage other authors to download the code, reuse or improve it for reproduction of these results or for similar analyses. 

\acknowledgements{
This publication has been made possible by the participation of more than 35,000 volunteers on the Milky Way Project. Their contributions are acknowledged individually at http://www.milkywayproject.org/authors. 

The Milky Way Project, and RJS were supported by The Leverhulme Trust. MSP was supported by a National Science Foundation Astronomy \& Astrophysics Postdoctoral Fellowship under award AST-0901646. Support for the work of KS was provided by NASA through Einstein Postdoctoral Fellowship grant number PF9-00069 issued by the Chandra X-ray Observatory Center, which is operated by the Smithsonian Astrophysical Observatory for and on behalf of NASA under contract NAS8-03060. GW-C gratefully acknowledges support from the Brinson Foundation grant in aid of astrophysics research at the Adler Planetarium.

SK is grateful to Neil Crighton and Ramin Skibba for assistance with clustering analyses and code checks; and to Mark Thompson and James Urquhart for advice regarding the RMS survey. Claudia Cyganowski, Kim Arvidsson, Chris Evans and Henrik Beuther provided useful comments on paper drafts.

This work is based on observations made with the Spitzer Space Telescope, which is operated by the Jet Propulsion Laboratory, California Institute of Technology under a contract with NASA. The Red MSX Source survey database (www.ast.leeds.ac.uk/RMS) was constructed with support from the Science and Technology Facilities Council of the UK. SK made extensive use of NASA's Astrophysics Data System Bibliographic Services for this work.}

\facility{
{\it Facilities:} \facility{Spitzer (IRAC, MIPS)}, \facility{MSX}
}



\bibliographystyle{apj}
\bibliography{mwp-triggering}

\end{document}